\documentclass{emulateapj}
\usepackage{apjfonts}

\newcommand{\msun}{\mbox{ M$_{\odot}$}}
\newcommand{\Zsun}{\mbox{ Z$_{\odot}$}}

\newcommand{\photflux}{\mbox{ photons cm$^{-2}$ s$^{-1}$ sr$^{-1}$}}

\newcommand{\cmden}{\mbox{ cm$^{-3}$}}

\newcommand{\ergs}{\mbox{ erg s$^{-1}$}}
\newcommand{\ergcmdens}{\mbox{ erg cm$^3$ s$^{-1}$}}

\newcommand{\Mpc}{\mbox{ Mpc}}
\newcommand{\kpc}{\mbox{ kpc}}

\newcommand{\hunits}{\mbox{ km s$^{-1}$ Mpc$^{-1}$}}
\newcommand{\kel}{\mbox{ K}}
\newcommand{\bq}{\begin{equation}}
\newcommand{\eq}{\end{equation}}

\begin{document}

\lefthead{THE COSMIC WEB IN LY$\alpha$}
\righthead{FURLANETTO ET AL.}

\title{Mapping the Cosmic Web with Ly$\alpha$ Emission}

\author{Steven R. Furlanetto\altaffilmark{1}, Joop
Schaye\altaffilmark{2}, Volker Springel\altaffilmark{3}, \& Lars
Hernquist\altaffilmark{4}}

\altaffiltext{1} {Mail Code 130-33; California Institute of
  Technology; Pasadena, CA 91125; sfurlane@tapir.caltech.edu}

\altaffiltext{2} {School of Natural Sciences, Institute for Advanced
Study, Einstein Drive, Princeton NJ 08540; schaye@ias.edu}

\altaffiltext{3} {Max-Planck-Institut f\"{u}r Astrophysik,
Karl-Schwarzschild-Strasse 1, 85740 Garching, Germany;
volker@mpa-garching.mpg.de }

\altaffiltext{4} {Harvard-Smithsonian Center for Astrophysics, 60
Garden St., Cambridge, MA 02138; lars@cfa.harvard.edu }

\begin{abstract}

We use a high-resolution cosmological simulation to predict the
distribution of \ion{H}{1} Ly$\alpha$ emission from the low-redshift
($z \la 0.5$) intergalactic medium (IGM).  Our simulation can be used
to reliably compute the emission from optically thin regions of the
IGM but not that of self-shielded gas.  We therefore consider several
models that bracket the expected emission from self-shielded regions.
Most galaxies are surrounded by extended ($\ga 10^2 \kpc$) ``coronae''
of optically thin gas with Ly$\alpha$ surface brightness close to the
expected background.  Most of these regions contain smaller cores of
dense, cool gas.  Unless self-shielded gas is able to cool to $T <
10^{4.1} \kel$, these cores are much brighter than the background.
The Ly$\alpha$ coronae represent ``cooling flows'' of IGM gas
accreting onto galaxies.  We also estimate the number of Ly$\alpha$
photons produced through the reprocessing of stellar ionizing
radiation in the interstellar medium of galaxies; while this mechanism
is responsible for the brightest Ly$\alpha$ emission, it occurs on
small physical scales and can be separated using high-resolution
observations.  In all cases, we find that Ly$\alpha$ emitters are
numerous (with a space density $\sim 0.1\, h^3 \Mpc^{-3}$) and closely
trace the filamentary structure of the IGM, providing a new way to map
gas inside the cosmic web.

\end{abstract}


\section{Introduction}
\label{intro}

One of the most striking images to emerge from cosmological
simulations is the ``cosmic web''.  Matter collapses into a web
of moderately overdense sheets and filaments, with galaxies and galaxy
clusters forming through continued collapse at their intersections.
This picture explains both the qualitative distribution of galaxies in
redshift surveys (e.g., \citealt{delapparent86}) and many of the
characteristics of the Ly$\alpha$ forest (see e.g. \citealt{rauch98}).
However, mapping the cosmic web remains a difficult venture.
Ly$\alpha$ forest spectra probe only one-dimensional lines of
sight, while galaxy redshift surveys find only the small fraction of
baryons embedded inside galaxies, offering an indirect picture
of the gaseous filaments.

In this \emph{Letter}, we use a cosmological simulation to show that
surveys of hydrogen Ly$\alpha$ ($\lambda$1216 \AA) emission can
provide a powerful method to map the cosmic web.  Existing analytic
estimates suggest that the surface brightness of dense portions of the
IGM can be substantial \citep{hogan87,gould96}.  Such regions
generally lie within filaments but outside of galaxies, so Ly$\alpha$
emission offers a more direct map of the gas distribution in filaments
than do galaxy surveys.  In fact, Ly$\alpha$ emission from galaxies
has been used to detect a filament at high redshift \citep{moller01};
here we focus on intergalactic emission from low redshifts, where the
distances and sky background are relatively small and where Ly$\alpha$
maps of the cosmic web can be compared to existing galaxy surveys.  We
also show that Ly$\alpha$-emitting gas constitutes an intrinsically
interesting phase: gas that is cooling around collapsed halos
\citep{haiman01-lya,fardal01}. The distribution of Ly$\alpha$ emission
therefore provides a window onto the growth of bound objects.

\section{Simulation and Analysis}
\label{method}

We perform our analysis using the G5 cosmological simulation of
\citet{springel03}.  This smoothed-particle hydrodynamics simulation
included a multiphase description of star formation which incorporates
a prescription for galactic winds \citep{springel-sf}.  The simulation
assumed a $\Lambda$CDM cosmology consistent with the most recent
cosmological observations (e.g., \citealt{spergel03}).  It has a box
size of $100\, h^{-1}$ comoving Mpc (where the Hubble constant
$H_0=100\, h \hunits$ and $h=0.7$ in the simulation), a spatial
resolution of $8 h^{-1}$ comoving kpc, and a baryonic particle mass of
$3.26 \times 10^8 h^{-1} \msun$.

\begin{figure}[t]
  \plotone{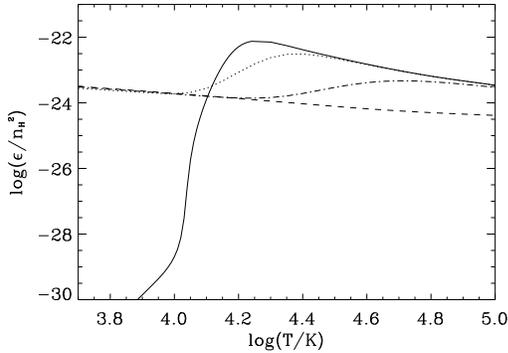}
  \caption{ The normalized Ly$\alpha$ emissivity
    $\varepsilon_\alpha/n_{\rm H}^2$, in units of $\ergcmdens$.  The
    solid curve includes only collisional processes.  The dashed,
    dot-dashed, and dotted curves allow photoionization and show the
    emissivity for gas with $n_{\rm H} = 10^{-6},\,10^{-4}$, and
    $10^{-2} \cmden$, respectively.
    \label{fig:emiss}}
\end{figure}

\citet[][{\ }hereafter F03]{furl03-metals}, who focus on UV emission
from metals in the IGM, describe the main components of our analysis
procedure; we briefly summarize them here.  We first compute two grids
of the Ly$\alpha$ emissivity $\varepsilon_\alpha$ as a function of
hydrogen density $n_{\rm H}$ and gas temperature $T$ using CLOUDY 96
(beta 5, \citealt{cloudy96}), assuming ionization equilibrium.  For
the first grid ($\varepsilon_\alpha^{\rm pi}$), we apply the
\citet{haardt01} ionizing background, which includes radiation from
galaxies and quasars (our main results are insensitive to the
background choice; F03).  For the second grid
($\varepsilon_\alpha^{\rm coll}$), we neglect all photoionization
processes; we use this grid to calculate the emissivity of
self-shielded gas.\footnote{Note that the electron density (and hence
$\varepsilon_\alpha^{\rm coll}$) becomes sensitive to the metallicity
for $T<10^4 \kel$.  However, this has negligible effects on our
results because at these temperatures the emissivity is vanishingly
small.}  We compare the two grids in Figure \ref{fig:emiss}. The solid
curve shows $\varepsilon_\alpha^{\rm coll}$ (note that
$\varepsilon_\alpha \propto n_{\rm H}^2$ when collisional processes
dominate), and the other curves show $\varepsilon_\alpha^{\rm pi}$ for
several different densities.  We see that $\varepsilon_\alpha^{\rm
coll}$ drops rapidly when the gas recombines at $T \approx 10^{4.2}
\kel$.  In contrast, $\varepsilon_\alpha^{\rm pi}$ is still large at
low temperatures because the gas remains photoionized.  As the density
and temperature increase, collisional processes become dominant and
$\varepsilon_\alpha^{\rm pi} \rightarrow \varepsilon_\alpha^{\rm
coll}$.

We wish to compute the emission from a slice of the simulation with
fixed $z$, $\Delta z$, and angular size.  We first randomly choose the
volume corresponding to this slice (see F03).  We classify particles
inside the slice as ``optically thin'' or ``self-shielded.''  The
latter set includes those dense and cool particles with a large enough
neutral fraction to shield themselves from the ionizing background.
We approximate self-shielding by assuming that all gas with $n_{\rm H}
> n_{\rm ss}$ and $T<T_{\rm ss}$ is optically thick.  We set fiducial
thresholds at $n_{\rm ss}=10^{-3} \cmden$ and $T_{\rm ss}=10^{4.5}
\kel$; radiative transfer calculations for self-gravitating clouds
like those described in \citet{schaye01-damp} find that self-shielding
becomes important at about this density, assuming that it remains cool
(see also \citealt{katz96}).  As shown in F03, most particles in this
regime lie on a well-defined curve in the $n_{\rm H}$--$T$ plane,
beginning at moderate overdensity and $T\sim 10^{4.25} \kel$ and
asymptotically approaching $10^4 \kel$ as the density increases to
$n_{\rm H} \sim 0.1 \cmden$.  This locus represents gas that is
cooling and falling into the centers of halos.  A small but important
fraction of particles (those that have recently been heated by shocks)
have slightly higher temperatures.  Once we have made this
classification, we assign emissivities to optically thin particles
using the $\varepsilon_\alpha^{\rm pi}$ grid and (in our fiducial
model) to self-shielded particles using the $\varepsilon_\alpha^{\rm
coll}$ grid.\footnote{Note that this is conservative in the sense that
we neglect photoionization by hard UV/X-ray background photons to
which the gas is still optically thin.}  Unfortunately, the emissivity
of self-shielded particles is rather uncertain because the simulation
does not include radiative transfer, metal line cooling, or local
ionizing sources and because we assume ionization equilibrium.  For
this reason, we examine their emission more closely in \S
\ref{selfshield}.

Active star formation also produces substantial Ly$\alpha$ emission
when gas in the host galaxy absorbs ionizing radiation from stars and
subsequently recombines.  In the simulation, star formation occurs in
any gas particle whose density exceeds $n_{\rm H}=0.129 \cmden$
\citep{springel-sf}.  For each such particle, we convert the star
formation rate (SFR) reported by the simulation to a Ly$\alpha$
luminosity via $L_\alpha = 10^{42} \, ({\rm SFR}/{\rm M}_\odot {\rm
yr}^{-1}) \ergs$, which is accurate to within a factor of a few for
metallicities between $0.05 \Zsun < Z < 2 \Zsun$ \citep{leitherer},
assuming that $\sim 2/3$ of ionizing photons are converted to
Ly$\alpha$ photons \citep{osterbrock89}.  The assumption of case-B
recombination should be valid provided that most ionizing photons are
absorbed in the dense interstellar medium of the galaxy.  The actual
Ly$\alpha$ luminosity of a galaxy depends strongly on the distribution
of ionizing sources, the escape fraction of ionizing photons, the
presence of dust, and the kinematic structure of the gas (e.g.,
\citealt{kunth03}), so this should be taken as no more than a
representative estimate.  
Note that any star formation unresolved by the simulation could
significantly change our results.  However, \citet{springel03} show
that the global SFR in the simulation matches observational estimates
well, so such difficulties are probably not severe.

Finally, we locate each particle on a pixelized map, convert its
emissivity to surface brightness, and smooth the result with a
Gaussian filter of full width at half maximum $\Delta \theta$ (F03).

\section{Results}
\label{results}

\begin{figure*}[t]
  \plotone{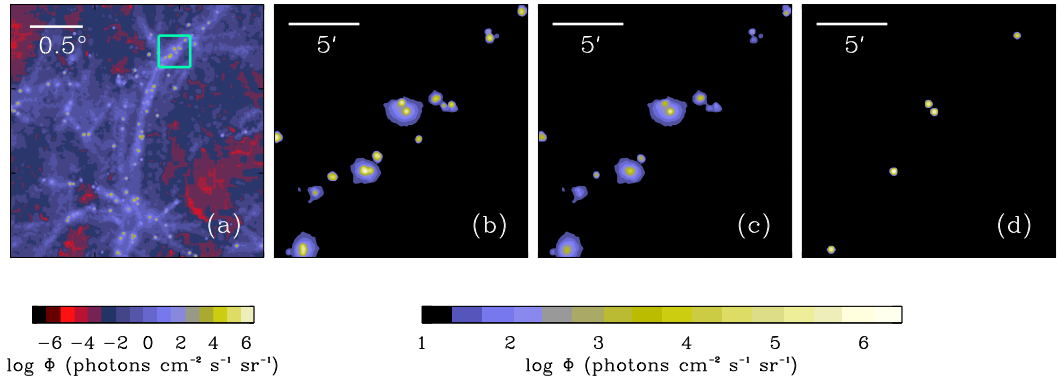}
  \caption{ Maps of Ly$\alpha$ surface brightness $\Phi $ for $z=0.15$
    and $\Delta z=10^{-3}$ ($\Delta \lambda=1.2$ \AA). Panel
    \emph{(a)} assumes an angular resolution of $\sim 29 \arcsec$; the
    rest show the region outlined in green with $7.2\arcsec$
    resolution (or $\sim 13\, h^{-1} \kpc$).  Except for panel
    \emph{(a)}, we exclude pixels with $\Phi < 10 \photflux$.
    \emph{(a), (b):} Fiducial model.  \emph{(c):}
    $\varepsilon_\alpha=0$ for self-shielded gas.  \emph{(d):}
    Estimated Ly$\alpha$ emission from star formation in the slice
    (such emission was not included in the other panels).
    \label{fig:lyamap}}
\end{figure*}

Figure \ref{fig:lyamap}\emph{a} shows a thin slice ($\Delta
z=10^{-3}$) of our simulated universe at $z=0.15$.  The panel shows a
large volume with coarse resolution; the colorscale includes the full
dynamic range of surface brightness $\Phi$.  Note how filaments stand
out clearly in the map.  Figure 2\emph{b} shows part of a filament
with a finer resolution ($\sim 13\,h^{-1}$ physical kpc).  We include
only those pixels with $\Phi > 10 \photflux$; this is a factor $\sim
10$ smaller than some observational \citep{brown00} and theoretical
\citep{haardt01} estimates of the diffuse background at $1200$--$2000$
\AA.  While the mean emission from filaments lies well below the
background, we find that numerous discrete regions trace their course:
the space density of these Ly$\alpha$ emitters is $\sim 0.1\, h^3
\Mpc^{-3}$.  Wide-field Ly$\alpha$ observations are thus a promising
technique to map out the cosmic web.  Note that decreasing the
threshold to $\Phi > 1 \photflux$ enlarges the coronae but does not
significantly increase their number density.

Ly$\alpha$ emission occurs over a wide range of physical scales:
generally, emitting regions have a central, high surface brightness
($\Phi \ga 10^3 \photflux$) core of size $\sim 30$--$50 h^{-1}
\kpc$ surrounded by an irregular, low surface brightness ``corona''
that can extend to scales $\ga 10^2 \kpc$.  Thus, the Ly$\alpha$
emission is much more extended than the galaxies, indicating that this
view of the cosmic web traces truly intergalactic gas.

\begin{figure*}[t]
  \plotone{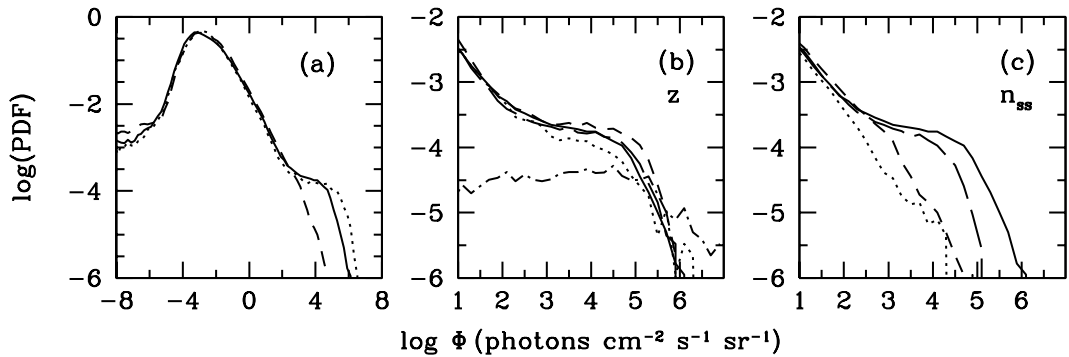}
  \caption{ The pixel surface brightness PDF in maps with $z=0.15$
    (unless otherwise specified), $\Delta z=10^{-3}$ ($\Delta
    \lambda=1.2$ \AA), and $13\, h^{-1} \kpc$ resolution.  In all
    panels, the solid curve shows our fiducial model.  \emph{(a):}
    $\varepsilon_\alpha=0$ for self-shielded gas (dashed curve) and no
    self-shielding cut (dotted curve).  \emph{(b):}
    $z=0.07,\,0.15,\,0.3$, and $0.45$ (dotted, solid, long-dashed, and
    short-dashed curves, respectively). Also shown is the Ly$\alpha$
    emission from star-forming gas (dot-dashed curve) at
    $z=0.15$.  \emph{(c):} $\varepsilon_\alpha=0$ for self-shielded
    gas, with $n_{\rm ss}=10^{-4},\,10^{-3}$, and $10^{-2} \cmden$
    (dotted, short-dashed, and long-dashed curves, respectively).
    \label{fig:basehist}}
\end{figure*}

Figure \ref{fig:basehist} shows histograms of the pixel flux
probability distribution function (PDF) ${\rm d}P/{\rm d} \log \Phi$.
The solid curve in each panel shows our fiducial model (using
$\varepsilon_\alpha^{\rm coll}$ for self-shielded gas).  Figure
3\emph{a} shows the PDF to very small $\Phi$; we see a broad peak
centered at $\Phi \sim 10^{-3} \photflux$ and a tail extending to much
higher surface brightness.  As described in Paper I, the location of
the peak is determined simply by ionization equilibrium in the
mean-density IGM and can be estimated analytically \citep{gould96}.
Note that the PDF depends strongly on $\Delta \theta$, because the
emitting regions have structure on small scales.

Figure 3\emph{b} shows that the PDF evolves very little from $z=0.07$
to $z=0.45$.  The fraction of bright pixels increases slowly with
redshift because both the density and temperature of the IGM increase
with redshift.

\subsection{Ly$\alpha$ Photons From Star Formation}
\label{sfgas}

Figure 2\emph{d} shows the Ly$\alpha$ emission from star formation.
While all star-forming regions are surrounded by intergalactic
Ly$\alpha$ emission, the converse is not true.  We do find that
Ly$\alpha$ emission nearly always surrounds star particles, but often
these are dwarf galaxies or lack ongoing star formation.  The
exceptions are galaxies inside hot clusters, which do not always host
Ly$\alpha$ emission.  Note that Ly$\alpha$ emission produced through
star formation is much more compact than the coronae (partly because
we assume that all the ionizing photons are absorbed locally).  Figure
3\emph{b} compares the mechanisms in a more quantitative form.  The
emission from star-forming gas in the simulation (dot-dashed curve)
dominates the brightest pixels but is unimportant for $\Phi \la 10^5
\photflux$.  If a large fraction of the ionizing photons escape their
host galaxy, the number of pixels affected by local star-formation
will increase, but their mean brightness will also decrease.  This
could be particularly important near active galactic nuclei (AGN),
which have large ionizing fluxes and large escape fractions.  AGN
could ``light up'' large volumes around their host galaxies in
Ly$\alpha$ emission \citep{haiman-rees}.  We note that the SFR, and
hence Ly$\alpha$ emission from star-forming gas, increases with
redshift \citep{hernquist03}.

\subsection{Self-Shielded Gas}
\label{selfshield}

The largest uncertainty in our calculation is probably the treatment
of self-shielded gas.  One problem is identifying which particles are
shielded from the ionizing background; we find, however, that our
results do not change significantly in the range $10^{-4} \cmden <
n_{\rm ss} < 10^{-2} \cmden$, because most gas of these densities has
$T \approx 10^{4.1} \kel$, where $\varepsilon_\alpha^{\rm pi} \approx
\varepsilon_\alpha^{\rm coll}$ (see Fig.~\ref{fig:emiss}).  A more
important concern is that we may have overestimated the temperature of
self-shielded particles because the simulation erroneously allows the
ionizing background to heat such particles and underestimates the
cooling rates in optically thick regions.  Radiative transfer
calculations with CLOUDY suggest that the simulation may have
overestimated the temperature of particles with $n_{\rm H} \sim
10^{-3} \cmden$ by $\ga 25\%$, which can be significant given the
steep dependence seen in Figure \ref{fig:emiss}.  The error increases
as the gas density and metallicity increase.  In a worst case
scenario, \emph{all} self-shielded gas would have $T < 10^{4.1} \kel$,
producing essentially no emission.  In Figure 2\emph{c}, we
approximate this case by setting $\varepsilon_\alpha=0$ for all
self-shielded gas.  Comparing to the fiducial model, we find that the
core brightness declines dramatically, but the extended coronae are
unaffected.  We conclude that the large-scale emission is due to
optically thin gas while the brightest regions consist of gas at or
near to the self-shielding threshold.  The resulting PDF is shown as
the short-dashed curve in Figure 3\emph{a}; it essentially eliminates
the bright tail of the distribution.  Figure 3\emph{c} shows that our
results \emph{do} depend on $n_{\rm ss}$ if self-shielded gas has
$\varepsilon_\alpha=0$.  The PDF changes little for $10^{-4} \cmden <
n_{\rm ss} < 10^{-3} \cmden$, because such gas has relatively low
density and hence small emissivity.  However, it approaches the
fiducial model if we set $n_{\rm ss} = 10^{-2} \cmden$, because gas
with $n_{\rm H} \ga 10^{-2} \cmden$ has $T \approx 10^{4.0} \kel$ in
the simulation and hence a small $\varepsilon_\alpha^{\rm coll}$.

On the other hand, the simulation also neglects local ionizing
sources.  We have already seen that most self-shielded particles
reside near galaxies; we would therefore expect the radiation field
around these particles to exceed the background.  In this case the
emissivity is limited by the ionizing flux reaching the gas.  We have
estimated the number of ionizing photons from young stars in \S
\ref{sfgas}, although we did not attempt to model the spatial
distribution of the resulting emission.  In addition, dense gas may no
longer be self-shielded, so the diffuse background will also
contribute to the Ly$\alpha$ emission.  We estimate this extra
component by assuming that no particles are self-shielded (dotted
curve in Figure 3\emph{a}).  The number of bright pixels
increases because $\varepsilon_\alpha^{\rm pi} \gg
\varepsilon_\alpha^{\rm coll}$ for $T < 10^{4.1} \kel$.
In either case, the net effect of local ionizing sources is to
\emph{increase} the Ly$\alpha$ emission.

\section{Discussion}
\label{discussion}

We study Ly$\alpha$ emission from gas in the IGM using a
high-resolution cosmological simulation.  We find that galaxies are
typically surrounded by ``coronae'' of Ly$\alpha$ emission with
bright, central cores (of diameter $\la 40\, h^{-1} \kpc$) surrounded
by much larger ($\ga 10^2\, h^{-1} \kpc$) regions with surface
brightness near the background.  While the number of Ly$\alpha$
photons due to star formation in the host galaxies can exceed that
emitted by the surrounding gas, the spatial extent of the star
formation is typically much smaller.  High angular-resolution
observations (probing physical scales $\la 15\, h^{-1}\kpc$) should be
able to separate the two components.  Detection of the bright cores is
feasible with existing technology, although the wide-field ultraviolet
spectrographs most useful for these studies have not yet been built.
The \emph{Galaxy Evolution Explorer} (\emph{GALEX}), with a large
field of view but relatively low spectral resolution ($\sim 10$ \AA),
may detect the brightest cores.  Furthermore, the strong correlation
between Ly$\alpha$ emission and galaxies may enable a statistical
detection of the signal in deep exposures.  A project underway to
construct a balloon-borne wide-field UV spectrograph will have a
limiting sensitivity $\Phi \sim 500 \photflux$ in a single night's
observation (D. Schiminovich, private communication).

We emphasize that the emission from self-shielded gas is uncertain,
because the simulation does not include all of the relevant physics.
We have considered a range of models bracketing these uncertainties
(see \S \ref{selfshield}): in our fiducial model, dense, cool gas
produces Ly$\alpha$ photons only through collisional processes, but we
also consider cases in which it has zero emissivity or in which we
allow photoionization.  The choice has little effect on the large-scale,
low surface brightness emission, which comes primarily from optically
thin gas, but it does strongly affect the luminous cores., where most
of the gas is shielded from the ionizing background.

We have also neglected dust, which can efficiently destroy Ly$\alpha$
photons.  Because the emitting region is often considerably larger
than the associated galaxy, much of the emitting gas may be relatively
pristine.  In the simulation, we find that the majority of particles
with large Ly$\alpha$ emissivity contain no metals (or dust), though
note that the simulation does not include mixing between simulation
particles.  At higher redshifts, several large-scale
Ly$\alpha$-emitting ``blobs'' have already been observed (e.g.,
\citealt{steidel00}), limiting the amount of dust in these
environments.

The structure and relatively large spatial size of Ly$\alpha$
emitters, together with their locations, suggest that they are
``cooling flows'' of gas onto halos.  Detailed, high resolution
Ly$\alpha$ observations therefore offer a means to study the accretion
and cooling of gas onto galaxies.  Galactic winds will also affect the
distribution of Ly$\alpha$ emission.  However, we find that the winds
in our simulations have only minor effects on the statistics of
Ly$\alpha$ emitters.

On larger scales, the Ly$\alpha$ emitters are distributed along
sheets and filaments.  Because these objects are
numerous (with a space density $\sim 0.1 h^3$ Mpc$^{-3}$),
Ly$\alpha$ line observations offer a new and powerful way to map the
cosmic web.  Figure \ref{fig:lyamap} shows that such observations do
not require high resolution; wide-field observations could efficiently
locate (in three dimensions) a large number of Ly$\alpha$ emitters.
In contrast to galaxy surveys, this approach reveals the location of
baryons \emph{outside} of galaxies.  Because Ly$\alpha$ emission
traces the cool gas accreting onto galaxies, it also selects a
different population of dark matter halos, including many with few (or
even no) associated stars.  Follow-up observations at high resolution
would allow us to compare Ly$\alpha$ emitters to galaxy surveys and
learn how galaxies accrete gas at the present day.

\acknowledgements We thank D. Schiminovich, C. Martin, and G. Ferland
for helpful discussions and B. Robertson for assistance with the
computational facilities.  This work was supported in part by The
W.M. Keck foundation, NSF grants PHY-0070928 and AST 00-71019 and NASA
ATP grant NAG5-12140.  The simulations were performed at the Center
for Parallel Astrophysical Computing at the Harvard-Smithsonian Center
for Astrophysics.

\end{document}